\documentstyle{mn}
\begin{document}

\def\simlt{\mathrel{\rlap{\lower 3pt\hbox{$\sim$}}
        \raise 2.0pt\hbox{$<$}}}

\def\simgt{\mathrel{\rlap{\lower 3pt\hbox{$\sim$}}
        \raise 2.0pt\hbox{$>$}}}
\title[First Stars Contribution to the Near 
Infrared Background Fluctuations] 
{First Stars Contribution to the Near
Infrared Background Fluctuations}

\author[M. Magliocchetti, R. Salvaterra, A. Ferrara]
{M. Magliocchetti, R. Salvaterra, A. Ferrara\\ 
SISSA/International School for Advanced Studies, Via Beirut 4, 34014 Trieste, Italy\\}

\maketitle
\vspace {7cm}

\begin{abstract}
We show that the emission from the first, metal-free stars inside Population III 
objects (PopIIIs) are needed to explain the level of fluctuations in the Near Infrared Background 
(NIRB) recently discovered by Kashlinsky et al. (2002), at least at the shortest 
wavelengths. Clustering of (unresolved) Pop IIIs can in fact 
account for the entire signal at almost all the 
$\sim$1-30 arcsec scales probed by observations in the J band. Their 
contribution fades away at  shorter frequencies and becomes 
negligible in the K band. ``Normal'', highly clustered, $\langle z\rangle\simeq 3$ 
galaxies undergoing intense star-formation such a those found in the Hubble 
Deep Fields can 'fill in' this gap and provide for the missing signal. 
It is in fact found that their contribution to the intensity fluctuations is 
the dominant one at $\lambda=2.17\mu$m, while it gradually looses importance 
in the H and J bands. The joint contribution from these two populations of
cosmic objects is able, within the 
errors, to reproduce the observed power spectrum in the {\it whole} 
Near Infrared range on small angular scales ($\theta\simlt 200^{\prime\prime}$ 
for Pop III protogalaxies). Signals on larger scales detected by other 
experiments instead require the presence of more local sources.

\end{abstract}

\begin{keywords}
galaxies: clustering - galaxies: infrared - cosmology: theory -
large-scale structure - cosmology: observations
\end{keywords}

\section{Introduction}
Different authors have reported the discovery of an excess in the Near 
Infrared Background (NIRB) which cannot be accounted for by normal galaxies
(see Hauser \& Dwek 2001 and references therein). This unaccounted excess
can be well fitted by the redshifted light of the first population of 
stars born inside the so-called Population III objects (Pop IIIs) (Santos, Bromm \&
 Kamionkowski 2001, Salvaterra \& Ferrara 2003 [SF03]).
More recently, Kashlinsky et al. (2002) have observed small-scale NIRB fluctuations
in the J, H and K bands, which cannot be due to local or low redshift
sources. 

The aim of this paper is to show that PopIIIs are indeed required to 
account for the observed small-scale ($\theta\simlt 200^{\prime\prime}$)
background fluctuations at least in the J band. Their contribution -- dominant 
for $\lambda=1.25\mu$m -- is found  
to rapidly decrease at longer wavelengths until it becomes negligible for 
$\lambda=2.17 \mu$m. We will also show that ``normal'', high ($\langle z \rangle\simeq 3$) 
redshift and highly clustered star-forming galaxies of the kind found in the 
Hubble Deep Fields can provide for the 'missing' power in the H and K bands, 
so that the joint contribution from these two populations can reproduce, 
within the errors, the observed power spectrum of intensity fluctuations 
at all NIR wavelengths. Note that, due to the high redshifts of the two candidate populations
($z\ga 8.8$, as found by SF03, for PopIIIs 
and $z\ga 2$ for ``normal'' star-forming galaxies), 
this is only true at small angular scales. Signals on intermediate-to-large 
scales detected by other experiments ($\theta\sim 0.7^\circ$; see e.g. 
Kashlinsky \& Odenwald 2000; Matsumoto et al. 2000) are instead expected to 
require populations of more local sources. 

The outline of the paper is as follows: in Sec. 2 we briefly describe 
the adopted model for the birth and evolution of Pop III protogalaxies. 
Sec. 3 presents our predictions for the intensity fluctuations of the 
NIRB due to the clustering of Pop III sources, while Sec. 4 compares such 
predictions with the available data and discusses the results. Finally, 
Sec. 5 summarizes our conclusions.

\section{The Model}

SF03 have shown that the NIRB data
(see e.g. Hauser \& Dwek (2001) and references therein) are well fitted by a 
model in which the contribution from PopIIIs is added to the ``normal'' 
galaxy background light as obtained from the {\it Hubble Deep Field} (Madau
\& Pozzetti 2000) or from the {\it Subaru Deep Field} (Totani et al. 2001).
We now briefly summarize some of the many aspects of the SF03 model.

The mean specific intensity of the background due to PopIIIs  
$I(\nu_{0},z_{0})$, as seen at a frequency $\nu_{0}$ by an observer at redshift 
$z_{0}$, can be written as
 
\begin{equation}
I(\nu_{0},z_{0})= \frac{1}{4\pi}\int^{z_{\rm max}}_{z_{0}}
\epsilon_\nu(z)e^{-\tau_{eff}(\nu_{0},z_{0},z)}\frac{dl}{dz}dz
\end{equation}
 
\noindent
(Peebles 1993). Here $\nu=\nu_0(1+z)/(1+z_0)$, $\tau_{eff}(\nu_{0},z_{0},z)$ is the
effective optical depth at $\nu_0$ of the IGM between redshift $z_0$ and
$z$, $dl/dz$ is the proper line element, and $z_{\rm max}$ is the redshift when 
the sources begin to shine; $\epsilon_\nu(z)$ is the 
comoving specific emissivity in units of  erg s$^{-1}$ Hz$^{-1}$
cm$^{-3}$, given by

\begin{equation}
\epsilon_\nu(z)=l_{\nu}(z)f_\star \frac{\Omega_b}{\Omega_M} \int^{\infty}_{M_{min}(z)}  
n(M,z) M  dM,
\end{equation}
\noindent
where $l_\nu(z)$ is the specific luminosity of the population at 
redshift $z$, computed using the spectra for metal-free stars obtained by
Schaerer (2002); $n(M,z)$ is the comoving number density of dark matter halos of mass
$M$ at redshift $z$ given by the Press \& Schechter formalism (Press \&
Schechter 1974). The integral
gives the dark matter mass per unit volume contained in dark matter halos with
mass greater than $M_{min}(z)$, where $M_{min}$ (computed by
Fuller \& Couchman 2000) is a cutoff mass below which halos
cannot form stars due to the lack of cooling. $\Omega_M$ and $\Omega_b$ are the total matter and 
baryon density\footnote{
We adopt the `concordance' model values for the cosmological parameters:
$h=0.7$, $\Omega_M=0.3$, $\Omega_{\Lambda}=0.7$, $\Omega_b=0.038$,
$\sigma_8=0.9$, and $\Gamma=0.21$, where $h$ is the dimensionless Hubble
constant, $H_0=100h$ km s$^{-1}$ Mpc$^{-1}$}
in units of the critical density $\rho_c=3H_0^2/8\pi G$.
The ratio $\Omega_b/\Omega_M$ converts the
integral into baryonic mass which is then turned into stars with an
efficiency $f_{\star}$.
This latter quantity is constrained by the contribution of the redshifted
light of PopIIIs to the NIRB (SF03). The PopIII star formation efficiency 
required to match the NIRB data depends essentially only on their 
initial mass function (IMF), being in the range 
$f_\star=10\%-50\%$ (the top-heaviest requiring lowest efficiency). 
The redshift at which the formation of metal-free stars ends is tightly 
constrained to be $z_{end}\simeq 8.8$ by the J band data (SF03).

\section{Contribution of clustering to background fluctuations}

The angular correlation function of intensity fluctuations $\delta I$ due to
inhomogeneities in the space distribution of unresolved sources
(i.e. with fluxes fainter than some threshold $S_d$) is defined as:
\begin{eqnarray}
C(\theta)=\langle \delta I(\theta^\prime, \phi^\prime)\; \delta
I(\theta'', \phi'')\rangle ,
\label{eq:ctheta}
\end{eqnarray}
where $(\theta^\prime, \phi^\prime)$ and $(\theta'', \phi'')$ identify
two positions on the sky separated by an angle $\theta$. The above expression 
can be written as the sum of two terms, $C_P$ and $C_C$, the first one due
to Poisson noise (i.e. fluctuations given by randomly distributed
objects), and the second one owing to source clustering (De Zotti et al. 1996 and 
Magliocchetti et al. 2001 for a detailed discussion). 
In the case of strongly 
clustered sources, as the NIR data seems to indicate 
(Kashlinsky et al. 2002), the Poisson term $C_P$ is found to be negligible 
when compared to $C_C$ (see e.g. Scott \& White 1999). We can therefore safely 
assume $C\simeq C_C$ (hereafter simply called $C(\theta)$), whose expression 
in the case of Pop III sources is given by
\begin{eqnarray}
C(\theta)=\left({1\over 4\pi}\right)^2  \int_{z_0}^{z_{\rm max}}\!\!\!\!\!\!\!\!\!\!\!\!dz\;
\frac{\epsilon_{\nu}^2(z)} {(1+z)^2}\;e^{-2\tau_{eff}}\left(\frac{dx}{dz}\right)
\nonumber \\
\times \int_{-\infty}^\infty du\;\xi_g(r,z), 
\label{eq:cth}
\end{eqnarray}
where $x=l(1+z)$ is the comoving radial coordinate, $r=(u^2+x^2\theta^2)^{1/2}$ 
(for a flat universe and in the small angle approximation), 
$\epsilon_{\nu}(z)$ the comoving specific emissivity at the frequency 
$\nu$, and $z_0$ and $z_{\rm max}$ are defined as in Sec. 2. 

The spatial two-point correlation function of a class of galaxies '$g$', 
$\xi_g(r,z)$, in general results from a complicated interplay between the 
clustering properties of the underlying dark matter and physical processes 
associated to the formation of such galaxies (see e.g. Peacock \& Smith 2000; 
Scoccimarro et al. 2001; Magliocchetti \& Porciani 2003). However, at 
high enough ($z\simgt 0.5-1$) redshifts, the probability for a dark matter 
halo to host more than one galaxy is negligible (see e.g. Somerville et al. 
2001), and one can write:
\begin{eqnarray}
\xi_g(r,z)=\xi(r,z)b^2_{\rm eff}(M_{\rm min},z),
\label{eq:xi}
\end{eqnarray}
where $\xi(r,z)$ is the mass-mass correlation function and $b^2_{\rm eff}
(M_{\rm min},z)$ the bias associated to all dark matter haloes massive enough 
to host a galaxy at redshift $z$.

In equation (\ref{eq:xi}), $\xi(r,z)$ has been obtained following the work by 
Peacock \& Dodds (1996; see also  Moscardini et al. 1998 and Magliocchetti et 
al. 2000), which provides an analytical way to derive the trend of the dark 
matter correlation function both in the linear and non-linear regime. Note 
that $\xi(r,z)$ only depends on the underlying cosmology and on the 
normalization of $\sigma_8$. The effective bias factor $b^2_{\rm eff}$ 
of all haloes with masses greater than some threshold $M_{\rm min}$ (quantity 
which generally may depend on the look-back time) is instead 
obtained by integrating the quantity $b(M,z)$ - representing the bias of
individual haloes of mass $M$ - opportunely weighted by their number
density $n(M,z)$: 
\begin{eqnarray}
b_{\rm eff}(z)=\frac{\int_{M_{\rm min}}^{\infty} dM\;b(M,z)\;n(M,z)}
{\int_{M_{\rm min}}^{\infty} dM\;n(M,z)},
\label{eq:beff}
\end{eqnarray}
where, in this case, the mass function is obtained according to the Press 
\& Schechter (1974) approach and we take the functional form for $b(M,z)$ from 
Jing (1998) (but see also Scannapieco \& Barkana 2002).

If we then plug into equation (\ref{eq:cth}) the expressions obtained by 
SF03 (see Sec. 2) for the specific emissivity 
$\epsilon_{\nu}(z)$, the halo cutoff mass  $M_{min}\equiv 
M_{min}(z)$, and use as the minimum redshift at which 
these sources contribute to the background the value $z_{end}=8.8$ 
(corresponding to the epoch at which metal-free star formation ends -- note that 
we are making the reasonable assumption that no such a source can be 
resolved by current observations), then, by also making use of expressions 
(\ref{eq:xi}) and (\ref{eq:beff}), we can derive predictions for 
the contribution of the clustering of (unresolved) PopIIIs to the 
background fluctuations at different wavelengths.    

\begin{figure}
\vspace{8cm} \includegraphics{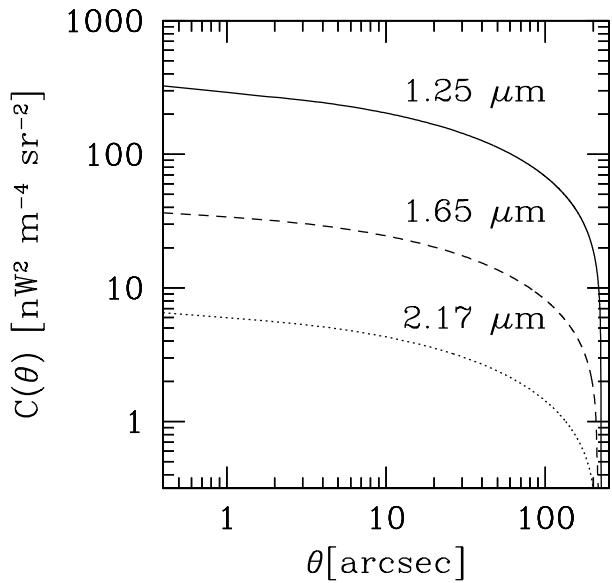}
\caption{Predictions for the correlation signal owing to intensity 
fluctuations of PopIII sources. The solid line represents the 
contribution in the J band, while dashed and dotted lines respectively 
illustrate the case for the H and K bands.   
\label{fig:cth}}
\end{figure}

The cases for $\lambda=1.25 \mu$m, $\lambda=1.65 \mu$m and 
$\lambda=2.17 \mu$m (respectively corresponding to J, H and K bands) are 
presented in Figure \ref{fig:cth} by the solid, dashed and dotted lines. 
Since these wavelengths at the redshifts under exam are always 
greater than the rest-frame Ly$\alpha$, the contribution from the $e^{-2\tau_{eff}}$
term in equation 4 can be neglected in the calculation of $C(\theta)$ (see SF03). 
 
The first feature to be noticed in the plot is the sharp drop 
of all the curves at $\theta\simeq 200$ arcsec. This is due to the fact that 
such an angular scale corresponds to distances $r\simgt 8.6$~Mpc (with the 
minimum value corresponding to $z_{end}=8.8$ in the adopted cosmology), 
where the spatial correlation 
function has already steepened from its power-law behaviour and rapidly 
approaches the zero value. As a first conclusion of this work we can 
then say that the clustering of unresolved PopIII sources cannot account for 
any of the observed fluctuations on scales $\simgt 200$ arcsec, which instead 
require much more local objects (for the Matsumoto et al. 2003 results the 
maximum acceptable redshift turns out to be $z\sim 1-2$).

The second point to stress is the remarkably different amplitude of 
the intensity fluctuations as evaluated at different frequencies. More 
specifically, $C(\theta)$ is found to decrease by about two orders 
of magnitude when going from $\lambda=1.25\;\mu$m to $\lambda=2.17\;\mu$m. 
As already argued by SF03, the reason for this 
decrement has to be found in the extremely strong Ly${\alpha}$ nebular 
emission line -- responsible for a considerable fraction of the PopIII emissivity -- which, 
once redshifted to the present time, gives its maximum contribution in the 
J band and rapidly disappears at the other two wavelengths under exam.    
We discuss the implications of these findings in the next Section.
 
\begin{figure}
\vspace{9cm} \includegraphics{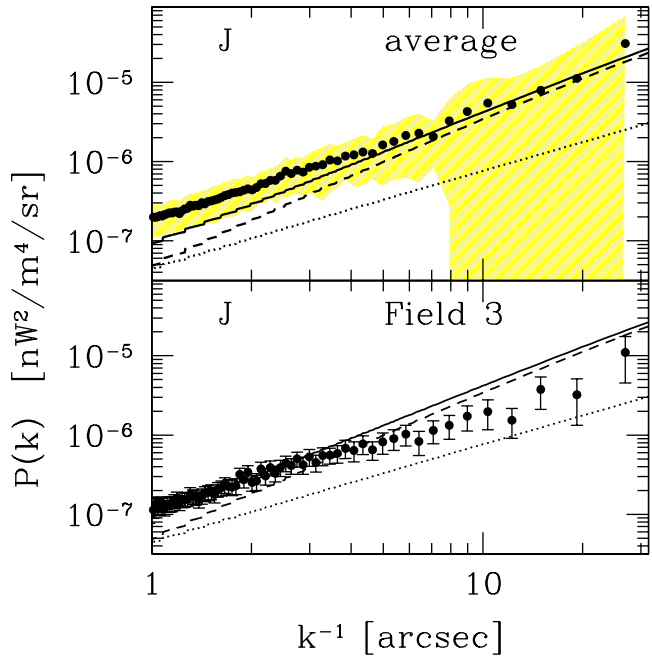}
\caption{Power spectrum of NIRB fluctuations at $\lambda=1.25\mu$m. 
The data are the results obtained by Kashlinsky et al. (2002) after atmospheric 
subtraction; the 
lower panel reproduces measurements from their Field 3 (considered by the 
authors as the most representative one), while the upper panel shows the 
signal obtained by averaging all the seven fields. The shaded area corresponds 
to uncertainties in the power spectrum arising from both statistical errors 
and field-to-field variations. In both panels, the dashed line represents 
the predicted contribution to NIRB fluctuations in the J band as due to 
clustering of PopIIIs, the dotted line that of ``normal'',  
highly-clustered, $\langle z \rangle\sim3$ star-forming galaxies and the solid line is the 
sum of both components (see text for details).
\label{fig:PkJ}}
\end{figure}

\section{Comparison with the data}
Kashlinsky et al. (2002) and Odenwald et al. (2003) have recently reported the first detection of 
small angular scale fluctuations in the Near Infrared Background. Their 
measurements were obtained by using long integration data constructed from 
2MASS (Two Micron Sky Survey) observations and by then coadding images 
in order to produce a $8.6^{\prime}\times 1^{\circ}$ field, divided into 
seven square patches 512$^{\prime\prime}$ on the side. In each patch, 
individual stars and galaxies were removed down to a magnitude limit which 
slightly varied from patch to patch, due to variations in the background level 
and associated noise.  The clipping algorithm was applied for a mash size of 3 by 3 pixels 
(corresponding to 9 square arcsecs). However, in order to test for its stability, the above procedure 
was also repeated with increasing mask sizes (up to 5 by 5 pixels) and the corresponding results were 
found in agreement with each other (S. Kashlinsky, private communication). Such an approach ensures the 
effective removal of entire galaxies at all redshifts but the lowest ones in which case more extended 
objects were identified by eye and removed by the clipping algorithm regardless. 
The remaining diffuse light was then Fourier 
transformed, with the resulting power spectrum showing a positive signal 
at all 1-30 arcsec scales and in all the seven patches.\\ 
The authors estimate contributions to the observed power spectrum from 
possible non-cosmological components such as atmospheric fluctuations, 
remaining Galactic stars, cirrus emission, zodiacal light (expected to be the largest 
source of uncertainty in the treatment of NIRB data), instrument noise 
and extinction, and conclude that they all have different slopes and 
negligible amplitudes when compared to the observed signal. Odenwald et al. 
(2003) and Kashlinsky et al. (2002) identify the measured power spectrum as 
due to clustering of faint (K $\ge$ 18.5-19 mag), high redshift, galaxies 
undergoing significant star-formation.

\begin{figure}
\vspace{9cm} \includegraphics{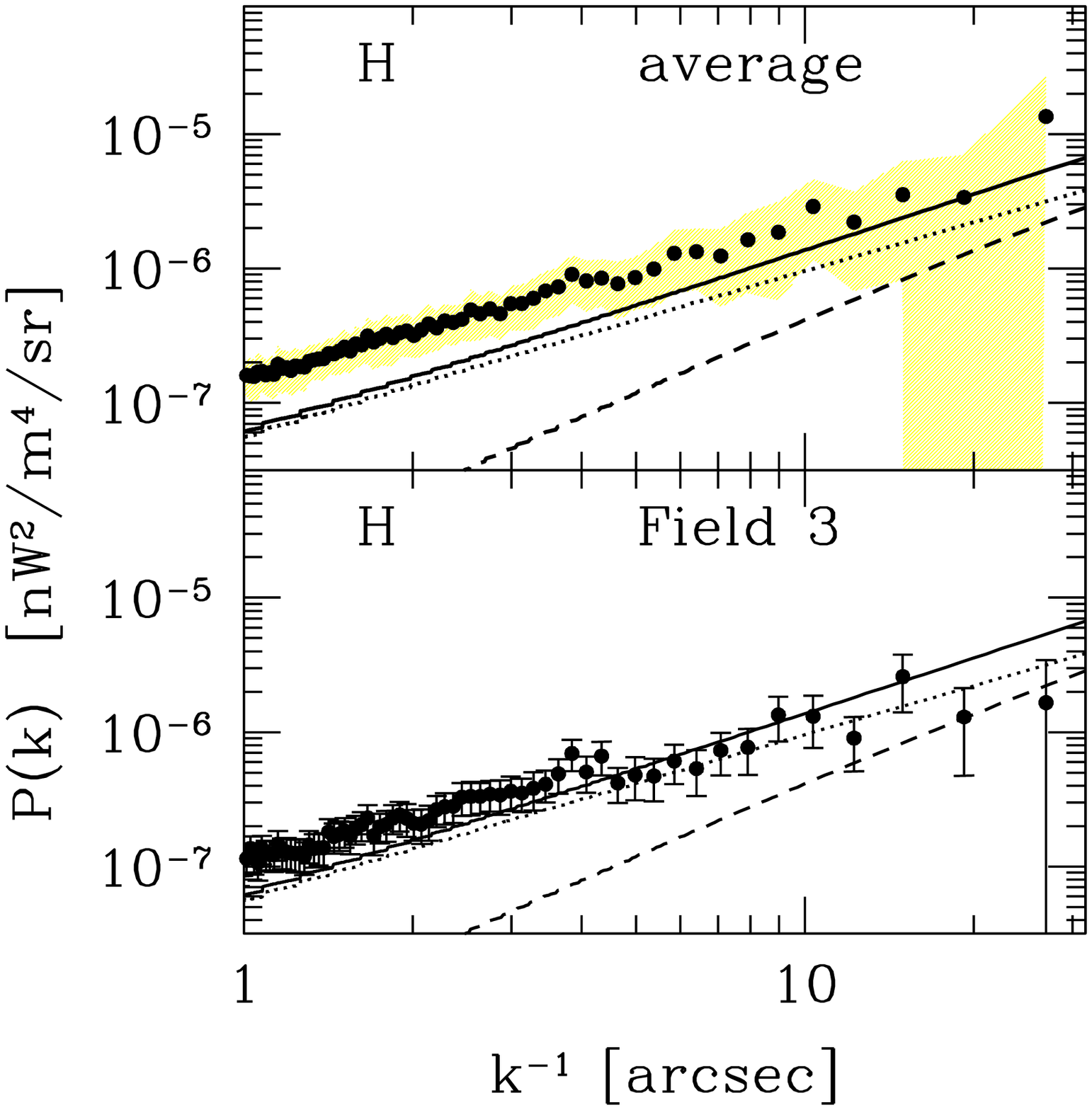}
\caption{As in Figure \ref{fig:PkJ}, but for $\lambda=1.65\mu$m.
\label{fig:PkH}}
\end{figure}

\begin{figure}
\vspace{9cm} \includegraphics{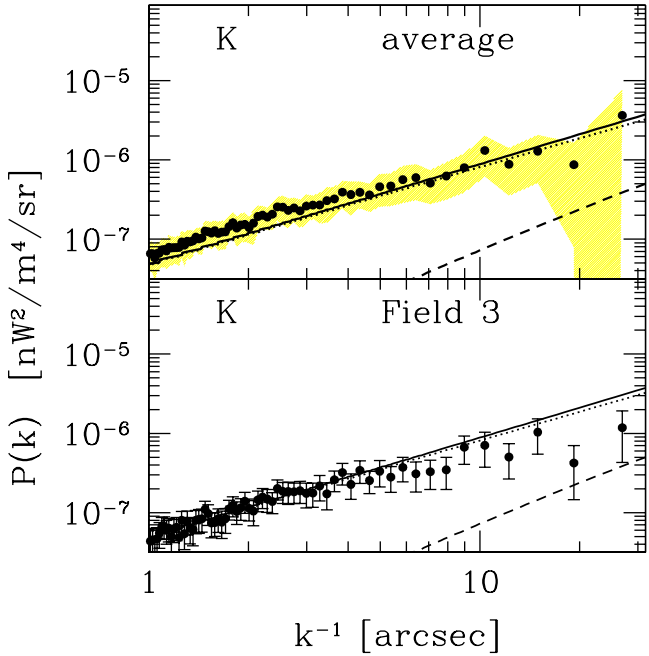}
\caption{As in Figure \ref{fig:PkJ}, but for $\lambda=2.17\mu$m.
\label{fig:PkK}}
\end{figure}

Their results (kindly provided by S. Kashlinsky in electronic form) are 
presented in Figures \ref{fig:PkJ}, \ref{fig:PkH} and \ref{fig:PkK} 
respectively for the J, H and K bands. The lower panels show the power 
spectrum as measured at the different wavelengths from Field 3, 
considered by Odenwald et al. (2003) as the most 
representative one due to its depth (limiting magnitudes for sources removal: 
J=20, H=19.2, K=19). The upper panels instead report the signal 
obtained by averaging measurements in all their seven fields. The shaded areas 
illustrate the levels of uncertainty due to both statistical (Poissonian) 
errors (corresponding to the errorbars in the lower panels) and 
field-to-field variations, found to be the dominant effect possibly because 
of the small area covered by each of the patches (cosmic variance). Note that 
the data has already been corrected for the atmospheric contribution. 

In order to compare these measurements with predictions from 
Sec. 3, we have converted our results for the intensity fluctuations 
$C(\theta)$ due to PopIIIs into the corresponding signal in the 
angular wavenumber $k$ space, $P(k)$, where 
\begin{eqnarray}
P(k)=2\pi\int_0^{\infty}C(\theta)\;J_0(k\theta)\;\theta\;d\theta,
\end{eqnarray}
with $J_0$ zero-th order Bessel function. 

The resulting power spectrum at the different wavelenghts is illustrated in 
Figures \ref{fig:PkJ}, \ref{fig:PkH} and \ref{fig:PkK} by the dashed lines. 
As it is clear by comparing data with model predictions in both panels of 
Figure \ref{fig:PkJ}, the clustering properties of PopIIIs can 
nicely explain the observed fluctuations in the J band, even though there 
seems to be some discrepancy at the smallest angular scales. The situation 
however gets increasingly worse when one goes to longer wavelengths and it 
is found that, while in the H band this class of sources can only (possibly) 
account for the observed spectrum at the highest angular scales probed by the 
data, their contribution in the K band falls short by at least one order of 
magnitude. Also, especially in the H and K bands, the predicted slopes for 
the power spectrum result too steep when compared to the measured ones.    

The missing ingredient to this analysis has to be searched amongst ``normal 
galaxies'' which can at least partially account for the extragalactic 
light emitted at the wavelengths under exam (see Madau \& Pozzetti 2000; 
Totani et al. 2001). In their work, Kashlinsky et al. (2002) and Odenwald et 
al. (2003) remove from the maps all resolved sources up to K$\simeq 19^m$ 
which, they claim, corresponds to the removal of possibly all resolvable 
galaxies to redshifts $z\simeq 1$ (with some allowance made for low-surface 
brightness systems). However, we argue that the population of galaxies 
responsible -- together with PopIIIs -- for the observed NIRB 
fluctuations must reside at redshifts $z\simgt 2$
\footnote {The possibility for a non-negligible part of the unaccounted signal to be due to residual 
disk emission from low-redshift galaxies whose nuclei have been removed can be discarded with a high 
confidence level both on the basis of the arguments introduced in this section re the clipping procedure 
and also by noticing 
that (relatively local) galaxies up to the clipping threshold K$\simeq 19^m$ contribute only a very small 
fraction of the CIB at this wavelength (see e.g. Kashlinsky et al. 2002). So, even if some (outer) parts 
of these galaxies were to remain, their contribution to the observed signal would be extremely small.}. 
The argument supporting this conclusion goes as 
follows.

The power spectrum of fluctuations as measured by 2MASS features a remarkable power-law 
behaviour down to angular scales of 1~arcsec. Under the hypothesis for this 
signal to be generated by large-scale structure, one has that -- since 
the correlation function of dark matter haloes (see equation \ref{eq:xi}) 
must drop 
to -1 on distances smaller than two virial radii due to spatial halo-halo 
exclusion (see e.g. Magliocchetti \& Porciani 2003) -- the power-law 
trend exhibited by the data down to 1 arcsec sets a strong limit on the 
minimum redshift at which these ``normal galaxies'' appear. 
More in detail, one has that for masses of the order of $10^8 M_{\sun}$ 
(value taken as a reasonable lower limit for the kind of objects under exam) and 
the cosmology adopted in this paper, the condition for a power-law behaviour 
to hold at the minimum scales probed by the data is only met for 
redshifts $z\simgt 2$.
 
What are then these $z\simgt 2$ ``normal galaxies'', faint enough not to be 
resolved by a K$\simlt 19$ survey, which can significantly contribute to the 
observed fluctuations of the NIRB, especially in the H and K bands? A 
natural candidate can be found in the population of high-$z$, star-forming 
galaxies detected in the Hubble Deep Fields to $I_{\rm AB}$=29. Magliocchetti 
\& Maddox (1999) have measured the clustering signal for this class of 
galaxies and shown that they are indeed strongly clustered, with an 
amplitude $A\simeq 7\times 10^{-3}$ (where the angular 
correlation function was parametrized as $w(\theta)=A\theta^{-0.8}$), constant within the 
errors at all redshifts $2\simlt z\simlt 4.8$. 

The contribution of this class of galaxies to intensity fluctuations in 
the NIRB can then be written as $C_{\rm gal}(\theta)\simeq I_{\rm B}^2
w(\theta)$, where $I_{\rm B}$ is the residual background intensity 
after source removal. Suitable values for $I_{\rm B}$ at the wavelengths 
relevant to this work can be obtained by taking the estimates for the 
total extragalactic light emitted in the J, H and K bands from Madau \& 
Pozzetti (2000), and by then subtracting the contributions from resolved 
sources as given in Kashlinsky et al. (2002). Note that the resulting 
$I_{\rm B}$'s, can in principle only be considered as upper limits, since the 
Kashlinsky and collaborators figures were given by assuming source removal 
up to $z\simeq 1$, while we are considering background fluctuations made 
by $z\simgt 2$ galaxies.   

The power spectrum of intensity fluctuations in the NIRB obtained for these 
``normal'' galaxies in the J, H and K bands is shown in Figures 
\ref{fig:PkJ}, \ref{fig:PkH} and \ref{fig:PkK} by the dotted lines. The solid 
lines instead illustrate the {\it total} signal steaming from the sum 
of the contribution of this class of sources with the one deriving from 
PopIIIs. ``Normal'', faint, high-redshift galaxies are able to 
reproduce the observed power spectrum at all angular scales in the K band, 
both in its amplitude and its slope. Their contribution however becomes 
less and less important as one moves towards shorter wavelengths and -- at 
$\lambda=1.25 \mu$m -- not only can just account for $\sim 10$\% of the 
observed signal, but also has a different (shallower) dependence on the 
angular scale than what the data indicates.\\
It is then found that, since PopIIIs and ``normal'' galaxies 
seem to be responsible for the 
observed fluctuations with a different relevance at the different wavelengths 
(PopIIIs mainly appearing in the J band, ``normal'' galaxies giving 
the strongest contribution in the K band), their sum is able to reproduce the 
data within the errors at {\it all} wavelengths (with some allowance made 
in the H band). In more detail, by considering 
both populations we can give account for the observed (roughly constant) 
amplitude of the observed power spectrum at the different wavelengths, and can 
also naturally explain the steepening of its slope when moving to shorter 
wavelengths as due to the intervening contribution of PopIII sources. 
Note that the match is particularly good, especially if one considers 
that {\it there are no free parameters in the model}. 
Our predictions do not depend on the exact functional form 
for the chosen IMF and star-formation efficiency $f_{\star}$, as long as their combination 
is able to reproduce the observed NIRB (SF03). Obviously, one of the main uncertainties 
of this work is associated to the determination of the slope of the predicted power spectrum. 
This is due to the lack of solid determinations of the two-point correlation function, from 
both an observational and a theoretical point of view, at the high redshifts and small masses 
under exam. The fact that our predictions result -- within the errors -- in agreement 
with the observed trend however gives us some confidence that the assumptions made in the course
of our analysis are (at least) plausible. 

As a last remark we note that, even if in the above analysis we have 
considered $I_{\rm B}$ in the different bands as an upper limit of the true 
background intensity, the fact that the data at $\lambda=2.17\mu$m is 
perfectly reproduced by the population of ``normal'' galaxies suggests 
this upper limit to be very close to the real value for otherwise the curves 
would have fallen below the observations. Under this assumption, there is 
then no need for the emergence of a third, still-unknown, population of 
objects, since PopIIIs, together with 
high-redshift/star-forming ``normal'' galaxies, can fully explain the 
observed level of fluctuations. 

 \section{conclusions}  
In this paper we have used the model of SF03
to derive the expected contribution of the clustering of 
(unresolved) PopIIIs 
to the intensity fluctuations of the NIRB. Comparisons with the recent 
observations by Odenwald et al. (2003) show that this class of sources 
can indeed account for the observed level of fluctuations at almost all the 
$\sim$1-30 arcsec scales probed by the data, at least in the J 
band where the Ly${\alpha}$ nebular 
emission line -- responsible for most of the PopIII emissivity -- 
once redshifted to the present time gives its maximum contribution. 
Their relevance however rapidly fades away for 
shorter frequencies and becomes negligible in the K band.

``Normal'', highly clustered, $2\simlt z \simlt 5$ 
galaxies undergoing intense star-formation such a those found in the Hubble 
Deep Fields can 'fill in' this gap and provide for the missing signal.
If we in fact assume these objects to produce a clustering signal 
as measured by Magliocchetti \& Maddox (1999) and associate them to a 
background intensity in in the J, H and K bands as taken from 
Madau \& Pozzetti (2000) (after source removal), we find that their 
contribution to the intensity fluctuations is 
the dominant one at $\lambda=2.17\mu$m, while it gradually looses importance 
in the H and J bands. 

The sum of the two PopIIIs and ``normal'' 
galaxies components is able, within the errors, to reproduce the observed 
power spectrum of intensity fluctuations in the {\it whole} 
Near Infrared range on small angular scales ($\theta\simlt 200^{\prime\prime}$ 
for PopIII sources). The match is particularly good, especially if 
one considers that our model is not associated to any free parameter. 
Signals on larger scales as detected by other 
experiments (see Hauser \& Dwek 2001 for review) are instead expected to 
require populations of more local sources.

The present results strongly support the scenario presented by 
Schneider et al. (2002) (see also Scannapieco, Schneider \& Ferrara 2003), 
in which the first stars where biased towards
high masses, thus producing copious amounts of ionizing photons and
heavy elements which caused cosmic reionization and enrichment
of the otherwise pristine intergalactic medium. It is only when 
the gas metallicity reached a critical value (still subject to uncertainties
due to the complicated physics of molecular cloud fragmentation) 
in the range $Z = 10^{-6}-10^{-4} Z_\odot$ that a relatively abrupt
transition to the `normal' star formation mode took place. 
This epoch has left a remarkable imprint both in the intensity and
the fluctuations of the NIRB that we have here isolated for the first time.

\noindent
\section*{ACKNOWLEDGMENTS}
Sasha Kashlinsky is warmly thanked for providing us with the data shown in
Figures 2, 3 and 4 and for useful clarifications on its derivation.


\begin{thebibliography}{}
\bibitem[1]{1} De Zotti G., Franceschini A., Toffolatti L., Mazzei P.,
Danese L., 1996, Ap. Lett.Comm., 35, 289
\bibitem[]{}Fuller T. M. \& Couchman H. M. P., 2000, ApJ, 544, 6
\bibitem[13]{13}Jing Y.P., 1998, ApJ, 503, L9
\bibitem[23]{23}
Hauser M., Dwek E., 2001, ARA\&A, 39, 249
\bibitem[18]{18} Kashlinsky A., Odenwald S., 2000, ApJ, 528, 74
\bibitem[3]{3} Kashlinsky A., Odenwald S., Mather J., Skrutskie M.F., Cutri 
R.M., 2002, ApJ, 579, L53
\bibitem [21]{21} Madau P., Pozzetti L., 2000, MNRAS, 312, L9
\bibitem[22]{22}
Magliocchetti M., Maddox S.J., 1999, MNRAS, 306, 988
\bibitem[10]{10}
Magliocchetti M., Bagla J., Maddox S.J., Lahav O., 2000, MNRAS, 314, 546
\bibitem[2]{2} Magliocchetti M., Moscardini L., Panuzzo P.,
Granato G.L. De Zotti G., Danese L. 2001, MNRAS, 325, 1553
\bibitem[7]{7} Magliocchetti M., Porciani C., 2003, MNRAS, submitted, astro-ph/0304003
\bibitem [19]{19} Matsumoto T. et al., 2000, in 'ISO Surveys of a Dusty 
Universe, ed. D. Lemke, M. Stickel \& K. Wilke, vol. 548, 96
\bibitem[15]{15} Matsumoto T. et al., 2003, ApJ, submitted
\bibitem[9]{9}
Moscardini L., Coles P., Lucchin F., Matarrese S., 1998, MNRAS, 299, 95
\bibitem[16]{16}
Odenwald S., Kashlinsky A., Mather J.C., Skrutskie M.F., Cutri R.M., 2003, 
ApJ, 583, 535
\bibitem[6]{6}Peacock J.A., Dodds S.J., 1996, MNRAS, 267, 1020
\bibitem[5]{5} Peacock J.A., Smith R.E., 2000, MNRAS, 318, 1144
\bibitem[44]{44}Peebles P. J. E., 1993, {\it Principles of Physical Cosmology}, 
Princeton Univ. Press, Princeton, NJ
\bibitem[12]{12} Press W.H., Schechter P., 1974, ApJ, 187, 425
\bibitem[14]{14} Salvaterra R., Ferrara A., 2003, MNRAS, in press, astro-ph/0210331 (SF03)
\bibitem[a]{a} Santos M. R., Bromm V., Kamionkowski M., 2002, MNRAS, 336, 1082
\bibitem [s]{s} Scannapieco E., Barkana R., 2002, ApJ, 571, 585
\bibitem[x]{x} Scannapieco, E., Schneider, R. \& Ferrara, A. 2003, MNRAS, in press, 
(astro-ph/0301628)
\bibitem[y]{y} Schaerer D., 2002, A\&A, 382, 28  
\bibitem[77]{77} Schneider, R. Ferrara, A., Natarajan, P. \& Omukai, K. 2002, ApJ, 571, 30
\bibitem[11]{11} Scoccimarro R., Sheth R.K., Hui L., Jain B., 2001, ApJ, 546, 20
\bibitem[4]{4} Scott D., White M., 1999, A\&A, 346, 1
\bibitem[8]{8}
Somerville R.S., Lemson G., Sigad Y., Dekel A., Kauffmann G., White S.D.M.,
2001, MNRAS, 320, 289
\bibitem [20]{20} Totani T., Yoshii Y., Iwamuro F., Maihara T., Motohara K., 
2001, ApJ, 550, L137
\end{thebibliography}
\end{document}